# Relaxation in Quantum Systems

Roumen Tsekov

Department of Physical Chemistry, University of Sofia, 1164 Sofia, Bulgaria

A new discrete model for energy relaxation of a quantum particle is described via a projection operator, causing the wave function collapse. Power laws for the evolution of the particle coordinate and momentum dispersions are derived. A new dissipative Schrödinger equation is proposed and solved for particular cases. A new dissipative Liouville equation is heuristically constructed.

In the recent years, the scientific interest to the dynamics of open quantum systems has grown [1]. The description of the quantum relaxation is not trivial since it involves specific interactions between the quantum system and its surrounding. Usually, it is accompanied by collapse of the wave function [2]. Let us consider one-dimensional motion of a quantum particle on the $x$-axis, where some scattering centers are distributed. The particle is not affected by any force during its movement between the scattering centers. For this reason, it is described by the standard Schrödinger equation for a free quantum particle with mass $m$

$$i\hbar \partial_t \psi = -\hbar^2 \partial_x^2 \psi / 2m \qquad (1)$$

For the sake of the further analysis, an auxiliary normalized Gaussian wave function is introduced $\phi_k(x) \equiv \exp(-x^2/4\sigma_k^2)/(2\pi\sigma_k^2)^{1/4}$, where $\sigma_k^2$ is the position dispersion. Since the Fourier image of $\phi_k$ is also Gaussian, the corresponding root mean square velocity reads $\eta_k = \hbar/2m\sigma_k$. If the initial condition of Eq. (1) is set to $\psi(x,0) = \phi_0(x)$, the solution the Schrödinger equation is

$$\psi(x,\tau_1) = \phi_1(x)\exp(im\alpha_1 x^2/2\hbar) \qquad (2)$$

where $\sigma_1^2 = \sigma_0^2 + \eta_0^2 \tau_1^2$. The physical meaning of the parameter $\alpha_1 \equiv \eta_0^2 \tau_1 / \sigma_1^2$ can be elucidated from the flow velocity $V \equiv i\hbar \partial_x \ln(\bar{\psi}/\psi)/2m$. It represents the gradient $\partial_x V$.

Suppose the quantum particle reaches a scattering center at time $\tau_1$. The interaction of the particle with the center is not described by Eq. (1). Our definition is that it consists in randomization of the wave function phase, resulting in termination of the systematic flow ($V = 0$). Hence, the particle losses its orientation and direction of motion. Mathematically, it is equivalent to application of a projection operator $\hat{P}\psi \equiv |\psi|$, which takes solely the modulus of the wave function. Since $\hat{P}$ preserves the probability density, the latter is always Gaussian. After such a wave function collapse, the quantum dynamics starts again from Eq. (1) with new initial condition $\hat{P}\psi(x, \tau_1) = \phi_1$. Note that the considered interaction is dissipative, since the root mean square velocity $\eta_1 < \eta_0$ decreases after the collision. The next solution $\psi(x, \tau_2) = \phi_2 \exp(im\alpha_2 x^2/2\hbar)$ of Eq. (1) possesses the following parameters $\sigma_2^2 = \sigma_1^2 + \eta_1^2 \tau_2^2$ and $\alpha_2 = \eta_1^2 \tau_2 / \sigma_2^2$. Considering now $\tau_2$ as the moment of the second impact of the particle with another screening center, one can apply again the projection operator $\hat{P}\psi(x, \tau_2) = \phi_2$, leading to $\eta_2 < \eta_1 < \eta_0$ as well. The procedure described above could continue further and the dispersion of the Gaussian wave packet after $N$ collisions will be given by

$$\sigma_N^2 = \sigma_0^2 + \sum_{k=1}^{N} \eta_{k-1}^2 \tau_k^2 \tag{3}$$

Without loss of generality one can assume firm localization of the particle in the beginning, which corresponds to $\sigma_0^2 = 0$.

As a first particular example, let us consider the case of equidistant collisions, where an analog of the mean free path in gasses $\lambda \equiv \eta_{k-1} \tau_k$ remains constant. In this case, Eq. (3) reduces to the linear law $\sigma_N^2 = \lambda^2 N$, well known from the description of independent random jumps [3]. However, the time between two consecutive collisions $\tau_{k+1} = \lambda / \eta_k = 2m\lambda \sigma_k / \hbar = (2m\lambda^2 / \hbar) k^{1/2}$ increases with the collision number $k$, since the particle becomes slower during the evolution due to the energy loss at the collisions. The overall duration of the whole process is a sum of the individual time intervals

$$t = \sum_{k=0}^{N-1} \tau_{k+1} = (2m\lambda^2/\hbar) \sum_{k=0}^{N-1} k^{1/2} \approx (4m\lambda^2/3\hbar) N^{3/2} = (4m\lambda^2/3\hbar)(\sigma_N/\lambda)^3 \qquad (4)$$

Inverting now this equation yields the evolution of the particle position and momentum dispersions

$$\sigma_x^2(t) \equiv \sigma_N^2 = (3\lambda\hbar t/4m)^{2/3} \qquad \sigma_p^2(t) \equiv m^2\eta_N^2 = (\hbar^2 m/6\lambda t)^{2/3} \qquad (5)$$

These power laws show increasing of the position dispersion and decreasing of the momentum dispersion in time. The only parameter of the surrounding is the mean free path $\lambda$. The number of collisions $N(t) = (3\hbar t/4m\lambda^2)^{2/3}$ increases also nonlinearly in time. The average time between collisions $2m\lambda^2/\hbar$ coincides with that from our previous estimate [4] but since the velocity of the particle slows permanently down, Eq. (5) differs from the law $\sigma_x^2 = \lambda(\hbar t/m)^{1/2}$ derived previously. This is also evident from the expression $b(t) \equiv 2m/\tau_{N+1} = 2m\eta_N/\lambda = \hbar/\lambda\sigma_x$ for the particle friction coefficient, which decreases during the evolution. Equations (5) are the solutions of the following energy balance in the continuous form

$$b\partial_t \sigma_x^2 = \hbar^2/2m\sigma_x^2 = 2\sigma_p^2/m \qquad (6)$$

The second example is more suitable for description of the quantum particle diffusion in condensed matter. In this case, the interaction of the particle with the surrounding is almost continuous and one can accept that the collision time $\tau_k \equiv \tau$ remains constant. To compensate partially the loss of energy, one can consider also a finite temperature $T$, when the environmental particles (i.e. the scattering centers) are moving fast as well. Therefore, the evolution of the position dispersion of the Gaussian wave packet will obey the following equation

$$\sigma_x^2(t+\tau) = \sigma_x^2(t) + \eta^2(t)\tau^2 \qquad (7)$$

where $\eta^2 = (\hbar/2m\sigma_x)^2 + k_B T/m$. Due to the large density of the scattering centers, the collision time is very small. For this reason, one can expand in series the left-hand side of Eq. (7) to get

$$m\partial_t^2 \sigma_x^2 + b\partial_t \sigma_x^2 = \hbar^2/2m\sigma_x^2 + 2k_B T = 2\sigma_p^2/m \qquad (8)$$

where the friction coefficient $b \equiv 2m/\tau$ is constant now. In the case of a strong friction, one can neglect the first inertial term in Eq. (8) to obtain the following simplified equation

$$\partial_t \sigma_x^2 = 2D(1+\lambda_T^2/\sigma_x^2) \qquad (9)$$

where $D \equiv k_B T/b$ is the Einstein diffusion constant and $\lambda_T \equiv \hbar/2(mk_B T)^{1/2}$ is the thermal de Broglie wave length. The direct integration of Eq. (9) yields a quantum generalization of the Einstein law [5]

$$\sigma_x^2 - \lambda_T^2 \ln(1+\sigma_x^2/\lambda_T^2) = 2Dt \qquad (10)$$

At large time Eq. (10) tends asymptotically to the classical Einstein law $\sigma_x^2 = 2Dt$, while at short time it reduces to $\sigma_x^2 = (\hbar/2m)(2\tau t)^{1/2}$. The latter expression is always true at zero temperature $T \equiv 0$, where the momentum dispersion decreases in time as $\sigma_p^2 = (m\hbar/2)/(2\tau t)^{1/2}$. Interestingly in this case, the dispersion of time $\sigma_t^2 \equiv m^2 \sigma_x^2/\sigma_p^2 = 2\tau t$ follows a simple Brownian motion, while the position dispersion is defined via a second Brownian motion $\sigma_x^2 = (\hbar/2m)\sigma_t$ with the Nelson universal diffusion constant $\hbar/2m$.

We are tempted to develop a continuous model, where the wave function satisfies the Schrödinger equation $i\hbar\partial_t \psi = \hat{H}\psi$ with a general Hamiltonian. Its unperturbed solution between two consecutive collisions reads $\psi(x,t+\tau) = \exp(-i\hat{H}\tau/\hbar)\psi(x,t)$. The previously used projection operator is not convenient due to the properties of the modulus function. One can apply, however, another projection operator $\hat{P}\psi \equiv \psi_{Re}$, which is defined to preserve the real part of the wave function. Note that in this case $\hat{P}$ diminishes the probability density, i.e. the scattering center can capture the quantum particle. Using the general solution and accounting for the collapses, the wave function evolves via the equation $\psi(x,t+\tau) = \cos(\hat{H}\tau/\hbar)\psi(x,t)$. Expanding the latter in series for short collision time $\tau$ yields the following dissipative differential equation

$$\partial_t^2 \psi + 2\partial_t \psi / \tau + \hat{H}^2 \psi / \hbar^2 = 0 \qquad (11)$$

Equation (11) reduces naturally to the Schrödinger equation at $\tau \to \infty$ due to the lack of collisions. The solution of Eq. (11) in the energy representation reads

$$\psi_E \propto \exp\{[(1 - E^2\tau^2/\hbar^2)^{1/2} - 1]\, t/\tau\} \qquad (12)$$

At low energy, it reduces to $\psi_E \propto \exp(-\tau t E^2 / 2\hbar^2)$ predicting a Gaussian energy distribution. The corresponding root mean square energy fluctuation $\sigma_E = \hbar/(2\tau t)^{1/2}$ correlates well to the previously obtained momentum dispersion $\sigma_p^2 = m\sigma_E/2$ and root mean square time fluctuation $\sigma_t = \hbar/\sigma_E$. According to Eq. (12) the particle tends to occupy lower energy states during the evolution and it drops at the end on the zero energy level when the particle energy is dissipated completely. At high energy, Eq. (12) reduces to the expression $\psi_E \propto \exp(-iEt/\hbar - t/\tau)$, which is in fact the solution of a dissipative Schrödinger equation $\partial_t \psi + i\hat{H}\psi/\hbar = -\psi/\tau$ due to the BGK approximation [6]. It implies an energy distribution independent of the particle energy. Another dissipative model [7] predicts the diffusive Schrödinger equation $\partial_t \psi + i\hat{H}\psi/\hbar = D\nabla^2 \psi$, being the mean field approximation of quantum state diffusion theory [8].

As already mentioned before, Eq. (11) reduces at short time $t < \tau$ to the Schrödinger equation $\partial_t \psi = -i\hat{H}\psi/\hbar$, while at large time $t > \tau$ one can neglect the first inertial term to obtain

$$\partial_t \psi = -\tau \hat{H}^2 \psi / 2\hbar^2 \qquad (13)$$

It is easy to solve this equation for a free particle and the wave function in the momentum representation reads $\psi_p \propto \exp(-\tau t p^4 / 8m^2\hbar^2)$. Thus, the dispersion $\sigma_p^2 = 2m\hbar\Gamma^2(3/4)/\pi(2\tau t)^{1/2}$ evolves in time as that derived before. Using the Fourier transformation one can calculate the wave function in the coordinate space as well

$$\psi = F([\,],[1/2,3/4], X^4/2) - 4\Gamma^2(3/4)(X^2/\pi)F([\,],[5/4,3/2], X^4/2) \tag{14}$$

where $F(\cdot,\cdot,\cdot)$ is a generalized hypergeometric function and $X \equiv x/2(\hbar^2\tau t/m^2)^{1/4}$. Its plot is presented in Fig. 1. As is seen, the dependence looks like Gaussian but the wave function possesses also additional extrema. Therefore, the probability density exhibits several maxima but the small ones are negligible as compared to the central one. The corresponding position dispersion $\sigma_x^2 = 3\hbar\Gamma^2(3/4)(2\tau t)^{1/2}/4\pi m$ evolves in time as that derived before and the Heisenberg relation is always satisfied.

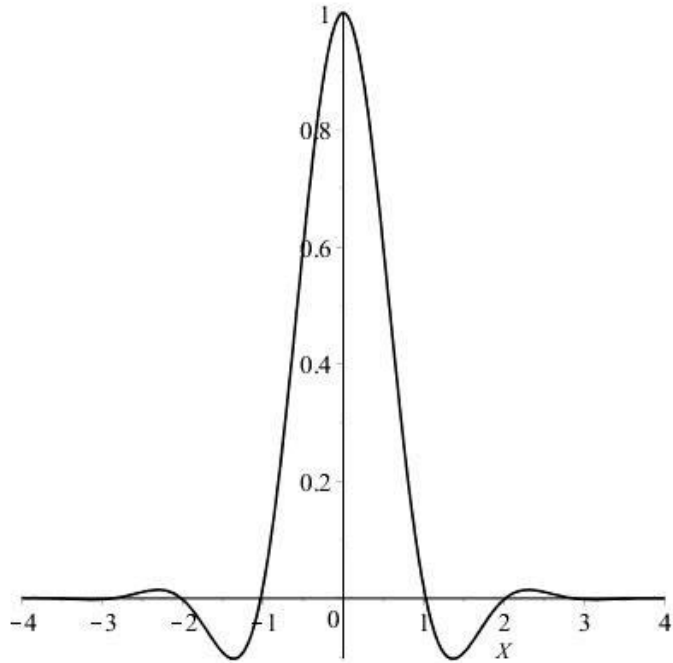

**Fig. 1.** The wave function $\psi$ in the coordinate representation

An interesting application of the model above is the description of a photon with the relativistic Hamiltonian $\hat{H} = c\hat{p}$, where $c$ is the speed of light in vacuum. In this case, Eq. (11) reduces to the dissipative wave equations

$$\partial_t^2\psi + 2\partial_t\psi/\tau - c^2\partial_x^2\psi = 0 \qquad \partial_t^2\psi_p + 2\partial_t\psi_p/\tau + (cp/\hbar)^2\psi_p = 0 \qquad (15)$$

The second equation here is the representation in the momentum space. If the collisions are very rare ($\tau \to \infty$) Eqs. (15) reduce to the well-known standard limits. In the more interesting case of frequent collisions with small $\tau$, the solutions of Eqs. (15) are Gaussian-like functions

$$\psi \propto \exp(-x^2/2c^2\tau t) \qquad \psi_p \propto \exp(-t\tau c^2 p^2/2\hbar^2) \qquad (16)$$

Surprisingly, the diffusion in the coordinate space obeys the Einstein law $\sigma_x^2 = c^2\tau t/2$, while the momentum dispersion $\sigma_p^2 = \hbar^2/2c^2\tau t$ reflects energy fluctuations $\sigma_E = c\sigma_p = \hbar/(2\tau t)^{1/2}$ as that obtained before.

The present dissipative model could be applied to classical systems as well. In the frames of the well-known BGK approximation, the relaxation of the probability density is given by [6]

$$\partial_t\rho + i\hat{L}\rho = (\rho_{eq} - \rho)/\tau \qquad (17)$$

where $i\hat{L}$ is the Liouville operator. In the case of lack of collision ($\tau \to \infty$) this equation reduces to the standard Liouville equation $\partial_t\rho + i\hat{L}\rho = 0$. A problem of Eq. (17) is that one should know in advance the equilibrium probability density $\rho_{eq}$. Applying a second derivative on time of the Liouville equation one can write in analogy of the present model the following new relaxation dynamic equation

$$\partial_t^2\rho + 2\partial_t\rho/\tau + \hat{L}^2\rho = 0 \qquad (18)$$

In the case of strong friction Eq. (18) simplifies further to

$$\partial_t\rho = -\tau\hat{L}^2\rho/2 \qquad (19)$$

The formal solution of this equation in operational form reads $\rho = \exp(-\tau \hat{L}^2 t/2)\rho_0$, where $\rho_0$ is the initial distribution density. The equilibrium distribution is the solution of the stationary Liouville equation $i\hat{L}\rho_{eq} = 0$.

To demonstrate the correctness of such modeling, let us consider a free particle. In this case, Eq. (19) reduces to a diffusion equation

$$\partial_t \rho = \tau p^2 \partial_x^2 \rho / 2m^2 \tag{20}$$

Its advantage is that the diffusion coefficient depends on the instant velocity of the particle. Using Fourier transformation along the particle coordinate simplifies the problem and the solution of Eq. (20) reads $\rho_q = \exp(-\tau t p^2 q^2 / 2m^2 - p^2/2mk_BT)/(2\pi m k_B T)^{1/2}$. The initial distribution corresponds to a Maxwell distribution in the momentum space $\exp(-p^2/2mk_BT)/(2\pi m k_B T)^{1/2}$ and a delta distribution $\delta(x)$ in the coordinate space. Integrating this result along the particle momentum yields the Fourier image of the particle concentration $c_q = 1/(1+2Dq^2 t)^{1/2}$, where the Einstein diffusion constant equals to $D = \tau k_B T / 2m$. It is possible to invert this image and the concentration profile acquired the analytical form $c = K_0(x/(2Dt)^{1/2})/\pi(2Dt)^{1/2}$, where $K_0(\cdot)$ is a modified Bessel function of second kind. The corresponding dispersion $\sigma_x^2 = 2Dt$ follows the Einstein law.

Let us conclude with the quantum aspects of the Virial theorem. According to the latter, the work done by the friction force is related to the particle kinetic energy via the relation $b\partial_t \sigma_x^2 = 2\sigma_p^2/m$, which is already derived in Eq. (6). To close this equation it is necessary to express the particle momentum dispersion. Generally, one can expand $\sigma_p^2$ in series of the Planck constant to obtain

$$\sigma_p^2 = \sum_{k=0}^{\infty} a_k \hbar^k \tag{21}$$

The first coefficient here is not quantum and one can recognize in $a_0 = mk_B T$ the thermal momentum dispersion. The second term is liner on $\hbar$ and, for this reason, it accounts for the quantum effects in the surrounding. Using the Heisenberg time-energy relation yields $a_1 = m/t$. The quadratic term is obviously due to the quantum nature of the particle and its coefficient equals to $a_2 = 1/4\sigma_x^2$ according to Heisenberg position-momentum relation. Hence, limiting the series in Eq. (21) up to the first three terms yields $\sigma_p^2 = mk_B T + \hbar m/t + \hbar^2/4\sigma_x^2$. Introducing it in the Virial theorem results in the following equation [4]

$$b\partial_t \sigma_x^2 = 2k_B T + 2\hbar/t + \hbar^2/2m\sigma_x^2 \qquad (22)$$

Comparing Eq. (22) with Eq. (8) shows that the liner $\hbar$-term is missing there, which is not surprising, since the previous analysis describes a quantum particle in classical environment. In the opposite case of a classical particle, moving in quantum environment at zero temperature, Eq. (22) provides the well-known result $\sigma_x^2 = (\hbar\tau/m)\ln(t/\tau)$ [9].